\documentclass[useAMS,usenatbib,asm]{mn2e}
 \usepackage{graphicx,fleqn,times}
\title[Star clusters associated with Sgr dSph galaxy]{
Old open clusters in the Sagittarius dSph tidal stream -- kith or kin?}
\author[G. Carraro and T. Bensby]{Giovanni Carraro$^{1}$\thanks{On 
leave from Dipartimento di Astronomia, 
Universit\'a di Padova, Italy}\thanks{E-mail:
gcarraro@eso.org (GC); tbensby@eso.org (TB)}
and 
Thomas Bensby$^{1}$\footnotemark[2]\\
$^{1}$European Southern Observatory, Alonso de Cordova 3107, 
Casilla 19001, Santiago 19, Chile}

\begin{document}

\date{Accepted 2009..... Received 2009...; in original form 2009...}

\pagerange{\pageref{firstpage}--\pageref{lastpage}} \pubyear{2009}

\maketitle

\label{firstpage}

\begin{abstract}
A widely supported formation scenario for the Galactic disc is that
it formed inside-out from material accumulated via accretion events.
The Sagittarius dwarf spheroidal galaxy (Sgr dSph) is the best example 
of a such accretion, and its ongoing disruption has resulted in 
that its stars are being deposited in the Milky Way halo and outer 
disc. It is therefore appealing to search for 
possible signatures of the Sgr dSph contribution to the build-up of the 
Galactic disc. Interestingly, models of the Sgr dSph stream indicate 
clearly that the trailing tail passes through the outer Galactic 
disc, at the same galactocentric distance as some anti-centre 
old open star clusters. We investigate in this Letter the possibility 
that the two outermost old open clusters, Berkeley~29 and Saurer~1, 
could have formed inside the Sgr dSph and then left behind 
in the outer Galactic disc as a result of tidal interaction with 
the Milky Way. The actual location of these two star clusters, 
inside the  Sgr dSph trailing
tail, is compatible with this scenario, and their chemical and 
kinematical properties, together with our present understanding 
of the age-metallicity relationship in the Sgr dSph, lends further 
support to this possible association. Hence, we find it 
likely that the old open star clusters Berkeley~29 and Saurer~1 
have extra-galactic origins.
\end{abstract}


\section{Introduction}

The Sagittarius dwarf spheroidal galaxy (hereafter simple referred to as the
Sgr dSph) is a nucleated 
dwarf galaxy on the verge to dissolve into
the Milky Way. Since its discovery in the 1990s (Ibata et al. 1994) 
an impressive amount of observational data have been collected to 
measure its properties, and significant theoretical efforts have 
been made to characterise its formation and evolution history. 
As a result, we now know that at the photometric centre of Sgr dSph
lies in M\,54 (NGC\,6715), a massive globular cluster showing 
multiple stellar populations (Siegel et al. 2007), which ended up in 
the centre of the Sgr dSph as a result of dynamical friction. 
Besides, the Sgr dSph
left behind an impressive star stream in the Galactic halo composed of a 
leading and a trailing tail (Newberg et al. 2002). This stream has been traced in detail by Majewski et al. (2004) with M giants from 2MASS survey using 
low-resolution spectroscopy. Additional information derive from 
spectroscopic studies which allowed to characterise the metallicity 
of the Sgr dSph centre and trailing stellar stream (Monaco et al. 2007). 
Important differences
have been found, with the stream being more metal-poor than the centre.
A family of star clusters have been found to be associated with the
Sgr dSph. Apart from M\,54, the other star clusters which are associated 
with the Sgr dSph are: Terzan\,7, Terzan\,8 and Arp\,2  
close to the core of the Sgr dSph (Ibata el al. 1994), and 
Palomar\,12 (Dinescu et al. 2000), Palomar\,2 (Majewski et al. 2004), and 
Whiting\,1 (Carraro et al. 2005, 2007a) in the stream. All of them are
globular clusters spanning ages from 7 to 12 Gyrs. Recently, 
Siegel et al.~(2007)  pointed out the presence of a much younger 
population associated to Sgr dSph, detected in the centre of the 
dwarf galaxy, close to M\,54. This implies that Sgr dSph has had 
protracted star formation episodes during its lifetime. 

The youngest globular clusters are on the other hand found in the stream, 
far from the central region (except for Terzan~8), and therefore one can 
expect to find traces of young stellar populations also outside the 
Sgr dSph centre.

In this Letter, we investigate the possibility that the two outermost 
old open star clusters in the Galactic disc, Berkeley~29 and Saurer~1,  
could have formed inside Sgr dSph, and then left behind in its trailing
tail due to tidal interactions with the Milky Way. These two star 
clusters appear to be twins, since they have very similar ages 
and metallicities. At the same time they bear significant differences 
with the bulk of old open clusters in the Galactic disc 
(Carraro et al.~2007b). First, as mentioned, they are currently the two 
most distant old open clusters that have been observed in the Galaxy;
secondly, they are located significantly above the Galactic plane 
(see Table~1). Furthermore, a detailed elemental abundance analysis 
of red giant stars in the two clusters by Carraro et al.~(2007a) revealed 
that they are more metal-poor than open star cluster of similar age in 
the solar vicinity. 

\begin{table*}
\tabcolsep 0.15truecm
\caption{Relevant parameters of the clusters under investigation.
Coordinates are for J2000.0 equinox. Coordinates in the Sgr orbital plane
are defined following Majewski et al. (2004). RA and DEC are also provided to
allow direct comparison with Mart\'inez-Delgado et al. (2004) models. }
\begin{tabular}{lccccccccccccccc}
\hline
\multicolumn{1}{c}{Cluster} &
\multicolumn{1}{c}{$l$} &
\multicolumn{1}{c}{$b$} &
\multicolumn{1}{c}{$RA$} &
\multicolumn{1}{c}{$DEC$} &
\multicolumn{1}{c}{$V_{\rm R}$} &
\multicolumn{1}{c}{$V_{\rm GSR}$} &
\multicolumn{1}{c}{Age} &
\multicolumn{1}{c}{$d_{\odot}$} &
\multicolumn{1}{c}{$d_{\rm GC}$} &
\multicolumn{1}{c}{$Z$} &
\multicolumn{1}{c}{$\lambda_{\odot}$} &
\multicolumn{1}{c}{$\beta_{\odot}$} &
\multicolumn{1}{c}{$\rm[Fe/H]$} &
\multicolumn{1}{c}{$\rm[Mg/Fe]$} &
\multicolumn{1}{c}{$\rm[Ca/Fe]$} \\
\noalign{\smallskip}
 & [deg] & [deg]& [deg]& [deg]& [km\,s$^{-1}$] & [km\,s$^{-1}$] & [Gyr] & [kpc] & [kpc] & [kpc] & [deg] & [deg] & & & \\
\hline
\smallskip
Berkeley~29   & 197.98 & +8.05 & 103.3 &  +1.9 &  24.6 &  -51.1 & 4.5 & 13.2 & 21.6 & 1.8 & 177.2 & 11.8 & -0.44 & -0.02 & +0.10 \\
Saurer~1      & 214.68 & +7.38 & 182.6 & +16.9 & 104.6 &  -32.7 & 5.0 & 13.2 & 19.2 & 1.7 & 181.8 & 27.8 & -0.38 & +0.03 & +0.18 \\
\hline
\end{tabular}
\end{table*}

While this might simply reflect the evidence for a Galactic radial 
abundance gradient (Magrini et al.~2009), we can not exclude the 
possibility that they formed outside the Galactic disc. In fact,
an association with the tidal debris of the Sgr dSph would 
strengthen the general idea that the Galactic disc formed inside-out 
from material deposited by accretion events. The outer metallicity 
plateau found for old open clusters in the Galactic disc 
(Magrini et al.~2009), would then be 
naturally explained, and would lend support to the models of 
Galactic chemical evolution proposed by Portinari \& Chiosi~(1999) 
and Hou et al.~(2000).

\begin{figure}
\includegraphics[width=\columnwidth]{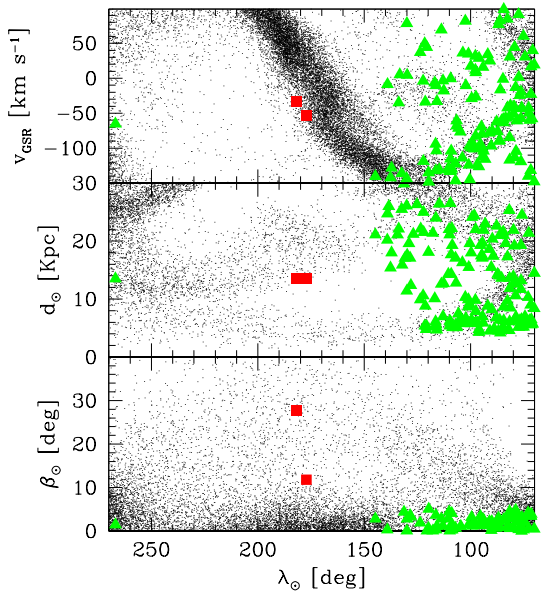}
\caption{
Location of Berkeley~29 and Saurer~1 (red squares). Bottom panel 
shows the position of the two clusters in the Sgr dSph orbital plane; 
middle panel their position as a function of heliocentric distance;  
upper panel their position as a function of galactocentric rest frame 
velocity. Dotted symbols are Sgr dSph points from Law et al.~(2005) 
model, which assumes a spherical halo. Green triangles are the 2MASS 
giants from Majewski et al.~(2004).
}
\end{figure}

\section{Kinematical signatures}

To probe the possible association of Berkeley~29 and Saurer~1 with 
the Sgr dSph, we will first consider how the position and velocity of 
these two star clusters compare with the Sgr dSph kinematics.

Law et al.~(2005) modelled the on-sky distribution and kinematics of 
2MASS M giant stars from the Majewski et al.~(2004) sample, to predict and 
map the spatial and kinematical properties of Sgr dSph trailing and 
leading tidal tails as a function of the shape 
(spherical, prolate or oblate) of the Galaxy dark matter (DM) halo. 
The N-body realisation for the favoured spherical DM halo
is depicted in the three panels of Fig.~1 for a suitable 
$\lambda_{\odot}$ range. $\lambda_{\odot}$, together with $\beta_{\odot}$, 
are longitude and latitude in a reference frame centred on Sgr dSph 
(the Sgr dSph orbital plane), defined  by Majewski et al.~(2004)\footnote{See 
also {\tt http://www.astro.virginia.edu/~smr4n/Sgr}} so that the results 
for Berkeley~29 and Saurer~1 could be directly compared with theirs for 
M giant stars and with the theoretical models of Law et al.~(2005). 
For comparison purposes, we also mark
the M giant stars observed by  Majewski et al.~(2004) (green triangles) in 
the same region of the Milky Way. In the Sgr dSph coordinate system, 
the main body of Sgr dSph is located at $\lambda_{\odot}=0^o$,
and $\lambda_{\odot}$ increases in the direction of the trailing stream.

The three panels show, from the bottom
to the top, the model predictions for $\beta_{\odot}$, the heliocentric distance $d_{\odot}$,
and Galactocentric rest frame velocity $V_{\rm GSR}$ of the debris 
particles, respectively. The same quantities are plotted for the M giant
stars as well. $V_{\rm GSR}$ have been calculated adopting the same
procedure and the choice of solar motion used by Majewski et al.~(2004).
Such plots have been routinely used both to select possible debris 
candidates for spectroscopic follow-up (Monaco et al.~2007), and for investigating and establishing the membership of star clusters to Sgr dSph
(NGC~5634: Bellazzini et al. 2002; Whiting~1: Carraro et al. 2007; AM~4: Carraro 2009). The two red squares indicate Berkeley~29 and Saurer~1. Unfortunately it is not possibly to compare the locations of these two 
clusters with M giant stars, since no M giant stars have been observed in this 
region of the sky. Nevertheless, the coordinates of Berkeley~29 and 
Saurer~1 (lower panel in Fig.~1) show that on the sky they lie in the direction of the trailing stream, although somewhat higher onto the 
orbital plane than the typical height of the closest M giant stars.
In the middle panel, we show that the heliocentric distance of the 
two clusters is compatible with the mean heliocentric distance of 
closest-to-the-Sun members of the same trailing stream. Although 
several M giant stars lie at comparable distances, no one unfortunately have 
been measured close to these Galactic directions. Finally, in the upper 
panel we plot the Galactocentric rest frame velocity $V_{\rm GSR}$ of 
Berkeley~29 and Saurer~1, and show that they are moving on similar orbits 
as the Sgr dSph trailing tail model particles.

We also investigate whether the estimates  for the distance, location and 
radial velocity of Berkeley~29 and Saurer~1 are consistent 
with the Martinez-Delgado et al.~(2004) and Helmi~(2004) model 
predictions, which are provided in Galactic coordinates and heliocentric radial velocity. A quick inspection of their plots indicates the same 
level of agreement we find with Law et al.~(2005) model.

These pieces of evidence suggest that it is
likely that both Berkeley~29 and Saurer~1 originated inside the 
Sgr dSph. In the next section we will look at the chemical properties 
of these two clusters, and compare them with our present understanding 
of the chemical evolution and star formation history of the Sgr dSph.

\begin{figure}
\includegraphics[width=\columnwidth,bb=18 185 569 530,clip]{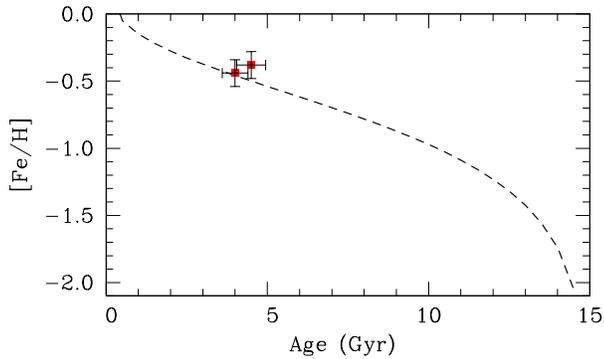}
\caption{The age-metallicity relationship for Sgr dSph from Layden \&
Sarajedini~(2000) is plotted as thick dashed line. The position of Berkeley~29
and Saurer~1 is indicated with red squares. Crosses refer to 
uncertainties in age
and metallicity.}
\end{figure}

\section{Chemical signatures}

A detailed elemental abundance analysis of the Berkeley~29 and Saurer~1
open star clusters have been
carried out by Carraro et al.~(2004) and were discussed in the general 
framework of the outer
Galactic disc chemical properties by Carraro et al.~(2007) and 
Magrini et al.~(2009).
As reported in Table~1, Berkeley~29 and Saurer~1 have similar 
ages and metallicities. At the age of these two clusters (4-5 Gyr)
the Galactic disc age-metallicity relation shows quite a large 
spread in metallicity (Carraro et al.~1998, Friel et al.~2002): 
from $\rm [Fe/H] \approx -0.5$ up to the solar metallicity of 
M~67 (NGC~2682). Such a spread is
larger than observational uncertainties  and is typically interpreted
as being caused by ancient merger events in the Galactic disc.
On the contrary, the age-metallicity relation of the Sgr dSph
appears to be better constrained (Forbes et al.~2004, Siegel et al.~2007).
The last determination by  Siegel et al. (2007) -which is based on
 Layden \& Sarajedini~2000 - is shown in Fig.~2, and 
it is consistent with a closed-box model and multiple bursts of
star formation over its entire lifetime (see Layden \& Sarajedini~2000 
for additional details). Berkeley~29 and Saurer~1  
possess the right combination of age and metallicity to fit perfectly 
into the age-metallicity relation of the Sgr dSph (see Fig.~2). 
The two clusters are probably part of an intermediate Sgr population 
(as defined by Siegel et al.~2000). This population has an average 
age between 4.5 and 6 Gyr, and a metallicity of
$\rm [Fe/H] \approx -0.5$.

Additional evidence of a possible kinship between the two open star
clusters and the Sgr dSph can be inferred from their 
detailed elemental compositions. Figure~3 shows the [Ca/Fe] and 
[Mg/Fe] abundance ratios as a function of [Fe/H] for 12 red giant 
stars in the Sgr dSph trailing tail analysed by Monaco et al.~(2007)
and another 27 stars from the Sgr dSph main body analysed by
Sbordone et al.~(2007). 
On top of the data for the Sgr dSph main body and trailing tail
in Fig.~3 we plot  the two open clusters from Carraro et al.~(2004). 
The error bars in the figure
represent a total error including 
the formal error in the mean abundance (i.e. line-to-line scatter 
divided by the square-root of the number of lines used), and
uncertainties in the abundances due to uncertainties in
the stellar parameters ($T_{\rm eff}$, $\log g$, and [Fe/H])
(all errors were added in quadrature).
Also in the figure we show the a sample of $\sim 700$ F and G dwarf
stars in the Solar neighbourhood by Bensby et al.~(2003, 2005) and
Bensby et al.~(in prep.).

From Fig.~3 it is evident that both the stars of the Sgr dSph main body
and trailing tail follow the same abundance trend, and that they 
with increasing [Fe/H] show a greater under-abundance
of $\alpha-$elements. The only difference is that the stream stars 
appear to be on average slightly more metal-poor than the main body stars.
We notice that the $\alpha-$element abundances of Berkeley~29 and Saurer~1 are
in good agreement with the Sgr dSph trailing tail as far as [Mg/Fe] 
is concerned. A slight discrepancy seems to exist for Ca. 
However, taking the uncertainties into account and the fact that
there might be slight systematic offsets between the data sets
due to different ways of normalising the abundances, we find
that also Ca is in reasonable agreement.
Comparing to the local disc stellar sample by Bensby et al.
we see that the two open clusters clearly have lower Mg abundances
but similar Ca abundances. 
A truly differential abundance analysis between all these stellar 
populations would be preferable, eliminating uncertainties arising from
different analysis methods that different studies use.

\begin{figure}
\includegraphics[width=\columnwidth,bb=18 160 592 718,clip]{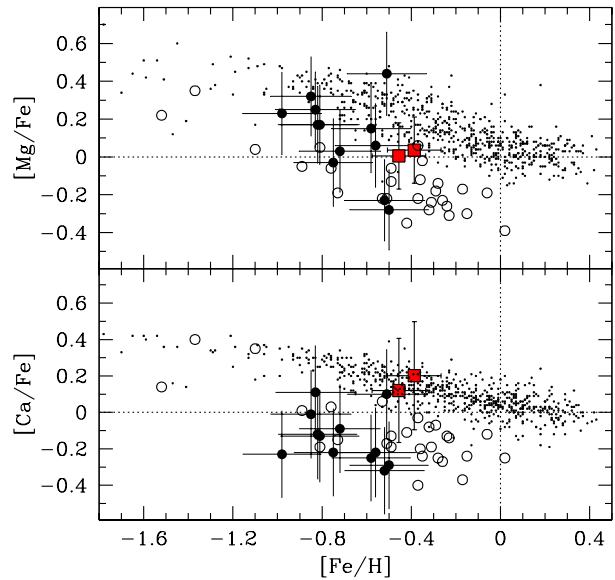}
\caption{[Mg/Fe] and [Ca/Fe] abundance ratios versus [Fe/H].
Filled circles mark the Sgr dSph trailing tail stars by
Monaco et al. (2007), open circles the Sgr dSph main body 
stars from Sbordone et al.~(2007), and the red squares the
Berkeley~29 and Saurer~1 open star clusters by
Carraro et al.~(2004). Error bars represent the total errors 
due to line-to-line scatter and uncertainties in the
adopted stellar parameters. Small dots mark the thin and thick 
disc stars from Bensby et al.~(2003, 2005, in prep.).}
\end{figure}

\section{Discussion and conclusions}

The formation and origin of the oldest
open star clusters in the outer Galactic disc have recently
been discussed (e.g., Frinchaboy et al. 2004; Carraro et al. 2007).
In fact, in the widely accepted scenario 
of inside-out formation of the Galactic disc (Magrini et al. 2009)
one would expect that the open star clusters of the 
outer Galactic disc  showed
signatures of extra-galactic origin.
Therefore, Frinchaboy et al. (2004) proposed that several 
globular clusters and old
open clusters may be associated with the Monoceros 
Ring (MRi), a gigantic star stream encompassing the entire Milky Way.
The MRi is believed to be tidal debris
left behind by the in-plane accretion  of a dwarf galaxy occurred 
4-8 Gyrs ago (Newberg et al. 2002).

Carraro et al. (2007) investigated in detail this possibility with the
available information on the kinematical and chemical properties
of the Monoceros Ring, 
together with model predictions by Pe\~narubia et al. (2005).
However, the poor knowledge of the Monoceros Ring properties
prevented any firm conclusions and led to a conservative
suggestion that Berkeley~29 and Saurer~1 were members of the Galactic disc.

In this Letter, we have compared the properties of Berkeley~29 and 
Saurer~1 with the Sgr dSph, whose chemical and kinematical 
properties are much better known (than the MRi). In particular, the 
trailing tail of Sgr dSph passes close to the Galactic disc periphery at the 
same distance as the two clusters.
We have shown that both kinematical and chemical properties of Berkeley~29
and Saurer~1 are compatible with a membership to the Sgr dSph.

\begin{figure}
\includegraphics[width=\columnwidth]{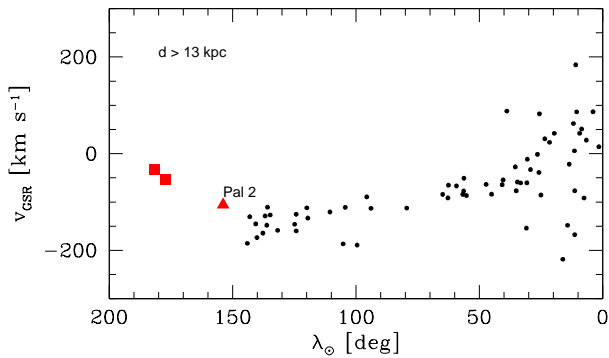}
\caption{The trailing tail of Sgr dSph galaxy as defined by M 
giant stars (Majewski et al. 2004). The big red squares mark the 
Berkeley~29 and Saurer~1, while the red triangle
indicates the position of the globular cluster Palomar~2}
\end{figure}

A possible weak point in the investigation could be
that since no M giant stars from the Majewski et al. (2004) survey
lie in the same region of the sky as the two open clusters, the
connection to the Sgr dSph had to be investigated by considering
model predictions from Law et al. (2005) only.
However, Majewski et al (2004) investigated the possibility that
several globular clusters might be associated to Sgr dSph.
As for Palomar~12, it lies very close to the trailing tail, as defined
by M giant stars, and additional chemical and kinematical evidence from
Cohen~(2004) support its association with the Sgr dSph.
Palomar~2, on the other end, falls in a region of the sky where
no M giant stars are present, as is the case for the two open cluster
Berkeley~29 and Saurer~1 in this study. However, Palomar~2 lies in a 
zone where the trailing tail, as defined by M giant stars,
likely  continues, and where such an extrapolation looks 
very reasonable, since Law et al. (2005) models perfectly
match the position of Palomar~2. It is therefore re-assuring
to find that Berkeley~29 and Saurer~1 happen to be situated in 
the same trailing tail extrapolation (see Fig.~4), lending further 
support to their association to the Sgr dSph.

Our suggestion that Berkeley~29 and Saurer~1 have formed inside Sgr 
dSph and then have been 
deposited in the outer Galactic disc about 5 Gyrs ago 
provides support to a scenario in which
the Galactic disc has grown  through repeated accretion events.
Additionally, this provides an explanation, several times advocated,
for the radial abundance gradient in the Galactic disc and its time
evolution (Carraro et al. 1998, Friel et al 2002, Magrini et al. 2009).
In this context, chemical evolution models employing an inside-out
formation mechanism for the disc (Portinari \& Chiosi) are clearly
favoured.

\section*{Acknowledgments}
We thank Lorenzo Monaco for providing the data used to build up Fig.~3
in electronic form and for useful discussions.

\end{document}